\documentclass[a4paper,11pt]{article}
\usepackage{pos}

\usepackage{graphicx}
\usepackage{amsmath}
\usepackage{epstopdf}
\usepackage[utf8]{inputenc}
\usepackage{pstricks}
\usepackage{color}

\usepackage[compat=1.0.0]{tikz-feynman}
\usepackage{simplewick}
\usepackage{color, colortbl}
\definecolor{Gray}{gray}{0.9}
\usepackage{float}
\usepackage{tikz-feynman}
\usepackage{feynmp}
\usepackage{forest}

\usepackage{verbatim}    
\usepackage{dcolumn}
\usepackage{bm}
\usepackage{epsfig}
\usepackage{braket}
\usepackage[normalem]{ulem} 
\usepackage{longtable} 

\def\beq{\begin{equation}}
\def\eeq{\end{equation}}
\def\bea{\begin{eqnarray}}
\def\eea{\end{eqnarray}}
\def\eq{\end{quote}}

\def\la{\langle}
\def\ra{\rangle}

\def\ba{\vspace*{-0.2cm}\begin{array}}
\def\ea{\end{array}\vspace*{-0.2cm}}

\def\als{\alpha_s}

\def\gg2{ \la\alpha_s G2 \ra}
\def\gg3{g^3f_{abc}\la G^aG^bG^c \ra}
\def\ggg4{\la\als^2G4\ra}

\def\beqa{\begin{eqnarray}}
\def\enqa{\end{eqnarray}}

\def\MeV{\nobreak\,\mbox{MeV}}
\def\GeV{\nobreak\,\mbox{GeV}}

\newcommand{\be}{\begin{equation}}
\newcommand{\ee}{\end{equation}}
\newcommand{\ben}{\begin{eqnarray}}
\newcommand{\een}{\end{eqnarray}}

\def\MeV{\mbox{ MeV}} 
\def\GeV{\mbox{ GeV}}

\title{Exotic hadrons in heavy ion collisions}

\author*[a]{Luciano Melo Abreu}

\affiliation[a]{Instituto de F\'isica, Universidade Federal da Bahia,\\
  Campus Universit\'ario de Ondina, 40170-115, Bahia, Brazil}

\emailAdd{luciano.abreu@ufba.br}

\abstract{
Nucleus-nucleus collisions offer a great opportunity for analyzing and determining the intrinsic nature of heavy and exotic hadrons.
In this sense, here we discuss how the production and dissociation of hadron states are affected by reactions during the expansion of hadronic matter in a heavy-ion collision environment. We give emphasis to recent works on exotic states, revisiting as a case study the time evolution of the abundances of the $X_{0,1}(2900)$ states. 
}

\FullConference{%
  XV International Workshop on Hadron Physics (XV Hadron Physics)
  13 -17 September 2021 \\
  Online, hosted by Instituto Tecnológico de Aeronáutica, São José dos Campos, Brazil\\
  }

\begin{document}
\maketitle

\section{Introduction}


Thanks to the joint efforts of experimentalists and theorists in recent decades, a great progress has been achieved in hadron physics. The constitution, production, decays and intrinsic properties of hadrons have been increasingly better understood~\cite{Tanabashi:2018oca,Nielsen,Brambilla:2010cs,Esposito:2014rxa,Chen,Hosaka,Brambilla:2019esw,Brodsky:2020vco}.  But despite this tremendous advancement, hadronic spectroscopy remains as a frontier field of study and one of the hot topics of particle physics, due to the contrasts still present between experimental observations and theoretical predictions. 
Below we mention one of the central issues that highlight this picture and is particularly of our interest.


The key fact is that new heavy hadrons that do not exhibit the expected properties of conventional hadrons have been observed from 2003 onwards by different collaborations (BELLE, BABAR, CLEO, BESIII, LHCb, etc.)~\cite{Tanabashi:2018oca}. Specifically, such states do not fit the usual description of quark models and cannot be classified as mesons or baryons, being called exotic states and usually denoted in the case of integer angular momentum as states $X, Y $ or $Z $~\cite{Tanabashi:2018oca,Nielsen,Brambilla:2010cs,Esposito:2014rxa,Chen,Hosaka,Brambilla:2019esw,Brodsky:2020vco}. Their underlying structures are still under debate, with several possibilities speculated: mesonic molecules (meson-meson bound states); tetraquarks (diquark-antidiquark couplings); glueballs (bound states of gluons); hybrids (quark-antiquark pair and a constituent gluon); manifestations of kinematic singularities present in processes whose intermediate particles are ``on-shell''- the so-called triangular singularities; among others~\cite{Brambilla:2019esw}.

As the structures mentioned above can be formed with the same quantum numbers, the community has endeavored to establish criteria to discriminate them. In this context, an interesting opportunity appears in the collisions of heavy ions at high energies, such as Au-Au in the Relativistic Heavy-Ion Collider (RHIC-BNL/USA)~\cite{star1,star2}, or Pb-Pb in the Large Hadron Collider (LHC-CERN/Switzerland-France)~\cite{alice1,alice2,alice3}.
In the initial stage of these collisions, nuclear matter is subjected to extreme conditions of densities, temperatures, intense external fields, etc., and experiences a phase transition to a locally thermalized state of deconfined quarks and gluons, the so-called quark-gluon plasma (QGP)~\cite{Brambilla,Roland,Armesto}.
After this initial stage, the QGP cools and hadronizes, thus forming a hadron gas, in which different types of hadrons interact inelastically and their respective abundances are modified. As time goes on, the system reaches chemical equilibrium, in which collisions become elastic, and abundances are fixed. In the end, at kinetic freeze-out temperature, the density becomes small, so there are no more interactions and the particles freely reach the detectors.

Due to the abundant number of heavy quarks and antiquarks produced in the initial stages of these collisions, exotic states can be produced in sufficient abundance to detect them experimentally, making possible their studies in these environments. Their initial abundances can be estimated via the quark coalescence model, whose advantage over the statistical model is the possibility of distinguishing the internal structure of the multiquark state as being a hadronic molecule, a compact tetraquark state, a kinematic effect or an excited state of a conventional meson~\cite{ExHIC,Cho2,Cho3,ChoLee1,XProd1,XProd2,XHMET,XHMET2,ZProd1,Abreu:2017pos,Abreu:2017cof,Abreu:2018ipp,Cleven:2019cre,Abreu:2018mnc,Abreu:2020ony,Abreu:2020jsl,LeRoux:2021adw,Abreu:2021mrq,Abreu:2021xpz,Llanes-Estrada:2021jud,Abreu:2021jwm,Llanes-Estrada:2021ffo}. The results suggest that the abundance of an exotic state such as $X(3872)$ is typically smaller in the compact multiquark interpretation than in the case of molecular configuration~\cite {XProd1,XProd2,XHMET,XHMET2,ZProd1}. Combining this data with the fact that only resonances with large natural widths are affected by subsequent hadronic evolution, it is possible to determine whether an exotic hadron produced in heavy ion collisions is a compact multiquark state or has a molecular configuration. Furthermore, it is worth mentioning that structures in the invariant mass spectrum of a specific decay channel, generated by kinematic effects, cannot be produced statistically, and thus will not appear in heavy ion collisions. Therefore, heavy ion collisions also allow the distinction of such effects from real resonances, as well as providing insight into the state structure.

In this sense, here we present some recent investigations on how the production and dissociation of exotic hadron states are affected by reactions during the expansion of hadronic matter in a heavy-ion collision environment. In particular, we take as one case study the estimation of the evolution of the abundances of the $X_J(2900)$ states. 
\section{The effective formalism}
\label{Effective Formalism}

Here we introduce the formalism employed to describe the time evolution of the abundance $N_X$ of a given state $X$ during the hadronic stage of heavy ion collisions. It is influenced by the interactions among $X$ and other light particles (denoted henceforth as $\phi$) constituting the medium, and can be investigated using the momentum-integrated evolution equation~\cite{ExHIC,Cho2,Cho3,ChoLee1,XProd1,XProd2,XHMET,XHMET2,ZProd1,Abreu:2017pos,Abreu:2017cof,Abreu:2018ipp,Cleven:2019cre,Abreu:2018mnc,Abreu:2020ony,Abreu:2020jsl,LeRoux:2021adw,MartinezTorres:2017eio}
\begin{eqnarray} 
\frac{ d N_{X} (\tau)}{d \tau} & = & \sum_{ a b; \phi  } 
\left[ \langle \sigma_{ a b  \rightarrow X \phi} 
v_{ a b} \rangle n_{a} (\tau) N_{b}(\tau)
- \langle \sigma_{ X \phi \rightarrow a b  } v_{X \phi } 
\rangle n_{\phi} (\tau) N_{X}(\tau) 
\right] 
\nonumber \\  
& & + 
\langle \sigma_{ a b \rightarrow X} 
v_{  a b } \rangle n_{ a } (\tau) N_{b}(\tau)
-  \langle \Gamma_{X \to a b } \rangle N_{X} (\tau)  , 
\label{rateeq}
\end{eqnarray}
where $n_{i} (\tau)$ and $N_{i}(\tau)$ represent the density and the abundance of a given particle in hadronic matter at proper time $\tau$; 
$\langle \sigma_{a b \rightarrow c d } v_{a b}\rangle $ is the thermally averaged cross section for the $X$ production and absorption reactions involving initial two-particles going into two final 
particles $ab \to cd$, and is given by
\ben
\langle \sigma_{a b \rightarrow c d } v_{a b}\rangle &  = & 
\frac{ \int  d^{3} \mathbf{p}_a  d^{3}
\mathbf{p}_b f_a(\mathbf{p}_a) f_b(\mathbf{p}_b) \sigma_{a b \rightarrow c d } 
\,\,v_{a b} }{ \int d^{3} \mathbf{p}_a  
d^{3} \mathbf{p}_b f_a(\mathbf{p}_a) f_b(\mathbf{p}_b) }
\nonumber \\
& = & \frac{1}{4 \beta_a ^2 K_{2}(\beta_a) \beta_b ^2 K_{2}(\beta_b) } 
\int _{z_0} ^{\infty } dz  K_{1}(z) \,\,\sigma (s=z^2 T^2) 
\nonumber \\
& & 
\times \left[ z^2 - (\beta_a + \beta_b)^2 \right]
\left[ z^2 - (\beta_a - \beta_b)^2 \right],
\label{thermavcs}
\een
with $\sigma_{a b \rightarrow c d }$ being the cross section for a process $a b \rightarrow c d $; $v_{ab}$ the relative velocity of the two initial  interacting 
particles $a$ and $b$; the function $f_i(\mathbf{p}_i)$ is the Bose-Einstein   
distribution of particles of species $i$, which depends on the temperature    
$T$; $\beta _i = m_i / T$, $z_0 = max(\beta_a + \beta_b,\beta_c 
+ \beta_d)$; and $K_1$ and $K_2$ the modified Bessel functions of second kind.

The second line of Eq.~(\ref{rateeq}) should be considered for those $X$ states having lifetime less than that of the hadronic stage (tipically of the order of 10 fm/c). In this case, the $X$ decay and its regeneration from the 
daughter particles  are therefore included, with the scattering cross section $\sigma_{ a b \rightarrow X}$ being given by the spin-averaged relativistic Breit-Wigner cross section~\cite{Abreu:2020ony}. 
In this sense, the thermally averaged decay width of $X$ is given by
$ \langle \Gamma_{X \to a b } \rangle =  \Gamma_{X \to a b } \left(m_{X} \right) 
 K_1  \left(m_{X} / T \right) / K_2  \left(m_{X} / T \right). $

We assume that the hadrons relevant in the reactions involving the $X $ state are in equilibrium, with the respective densities $ n_{i} (\tau)$
written in Boltzmann approximation~\cite{ChoLee1,XProd2,Abreu:2020ony,Koch}.
\ben n_{i} (\tau) &  \approx & \frac{1}{2 \pi^2}\gamma_{i} g_{i} m_{i}^2 
T(\tau)K_{2}\left(\frac{m_{i} }{T(\tau)}\right), 
\label{densities}
\een
where $\gamma _i$ and $g_i$ are the fugacity factor and the  degeneracy factor of the particle, respectively. 
The multiplicity $N_i (\tau)$ is easily obtained by multiplying $n_i(\tau)$ by the volume $V(\tau)$. In this approach the time dependence of $n_{i} (\tau)$ is encoded in the expressions for the temperature $T(\tau)$ and volume $V(\tau)$ used to model the dynamics of relativistic heavy ion collisions after the end of the QGP phase. We adopt the Bjorken picture with an accelerated transverse expansion, in which the 
hydrodynamical expansion and cooling of the hadron gas is based on the following parametrization~\cite{ChoLee1,XProd2,Abreu:2020ony}
\ben
T(\tau) & = & T_C - \left( T_H - T_F \right) \left( \frac{\tau - \tau _H }
{\tau _F -  \tau _H}\right)^{\frac{4}{5}} , \nonumber \\V(\tau) & = & \pi \left[ R_C + v_C 
\left(\tau - \tau_C\right) + \frac{a_C}{2} \left(\tau - \tau_C\right)^2 
\right]^2 \tau \, c , 
\label{TempVol}
\een
where $R_C $ and $\tau_C$  denote the final transverse  and longitudinal sizes of the QGP; $v_C $ and  $a_C $ are its transverse flow velocity and transverse acceleration at $\tau_C $; $T_C$ is the critical temperature for the QGP to hadronic matter transition; $T_H $  is the temperature of the hadronic matter at the end of the mixed phase, occurring at the time $\tau_H $; and the kinetic freeze-out temperature  $T_F $ leads to a freeze-out time $\tau _F $. As an example, in Table~\ref{param} these parameters are given  taking the scenario of central Pb-Pb collisions at $\sqrt{s_{NN}} = 5.02$ TeV at the LHC.

\begin{table}[h!]
\caption{Set of parameters used in Eq.~(\ref{TempVol}) for the hydrodynamical expansion in the scenario of central Pb-Pb collisions at $\sqrt{s_{NN}} = 5$ TeV at the LHC~\cite{ExHIC,Abreu:2020ony}. }
\label{param}
\begin{center}
\begin{tabular}{ c c c }
\hline
\hline
 $v_C$ (c) & $a_C$ (c$^2$/fm) & $R_C$ (fm)   \\   
0.5 & 0.09 & 11  
\\  
\hline
 $\tau_C$ (fm/c) & $\tau_H$ (fm/c)  &  $\tau_F$ (fm/c)  \\   
7.1  & 10.2 & 21.5
\\  
\hline
  $T_C (\MeV)$  & $T_H (\MeV)$ & $T_H (\MeV)$ \\   
 156 & 156 & 115   \\  
\hline
\hline
\end{tabular}
\end{center}
\end{table}


%
%
%
%
%
%
%
%
%
%
%
%
%
%
%
%
%
%
%


Hence, the point here is that the production and dissociation of hadron states might be influenced by reactions during the expansion of hadronic matter in a heavy-ion collision environment. In some cases, as the recently observed exotic hadronic states, this analysis might shed some light on their intrinsic structure. 
Their multiplicities can be estimated by solving Eq.~(\ref{rateeq}), with initial conditions given according to the coalescence model. In this model the yield of a hadron is calculated from the overlap of the density matrix of the constituents in an emission source with the Wigner function of the produced particle~\cite{ChoLee1}. Therefore, the information on the internal structure (as angular momentum, multiplicity of quarks, etc.) of a given state is contemplated. To illustrate this argument, in the next section we discuss a case study: the time evolution of the abundances of the $X_{0,1}(2900)$ states.

\section{A case study: the $X_{0,1}(2900)$ states}
\label{case}

Here we revisit a particular and interesting case firstly investigated in Ref.~\cite{Abreu:2020ony}: the time evolution of the abundance of the $X_{0,1}(2900)$ states. 
They have been reported recently by the LHCb collaboration, from the observation of an exotic peak in the $D^- K^+ $ invariant mass spectrum of the $B^+ \to D^+ D^- K^+ $ decay~\cite{LHCb:2020bls,LHCb:2020pxc}, which has been fitted to two resonances $X_{0,1}(2900)$ with the corresponding quantum numbers, masses, and widths:
\ben
J^P = 0^+ : \,\, M = (2866\pm 7) \MeV, \,\, \Gamma = (57\pm 13) \MeV ; \nonumber \\
J^P = 1^- : \,\, M = (2904\pm 5) \MeV, \,\, \Gamma = (110 \pm 12) \MeV  .  
\label{Xstates}
\een
Due to their minimum valence quark contents of four different flavors, i.e. $\bar{c}\bar{s} u d $, they were considered as the first-observed  fully-open charm tetraquarks.
A heated debate has taken place regarding their internal structure, with emphasis on the compact tetraquark interpretation, resulting from the binding of a diquark and an antidiquark; the meson molecule picture, based on bound states of spin-1 charmed $D_{(1)}^{(\ast)}$ and $ K^{(*)}$ mesons; kinematic effects caused by triangle singularities; and so on (see Ref.~\cite{Abreu:2020ony} for a more detailed discussion).

To contribute on this discussion of the discrimination of the $X_{0,1}(2900)$ nature, the analysis of their multiplicity in heavy-ion collisions appears as a promising approach~\cite{Abreu:2020ony}.       
Accordingly, considering the $X_J(2900)$ as a tetraquark state produced via quark coalescence mechanism from  the QGP phase at the critical temperature $T_c$ when the volume is $V_C$, then for $J=0$ it is a $S$-wave and for $J=1$ a $P$-wave. Their yields in these different situations are given in Table~\ref{Tab3}, assuming they are isoscalar states. For completeness, we also show in this Table the yields of $X_J(2900)$ as a weakly bound hadronic molecule from the hadron coalescence at $T_F$ and $V_F$, since they are dominantly formed at the end of the hadronic phase. In this context, the case $J=0$ is a $S$-wave hadronic molecule, while $J=1$ a $P$-wave.

\begin{table}[h!]
\caption{The $X_{0,1}(2900)$ yields in central Pb-Pb collisions at $\sqrt{s_{NN}} = 5.02$ TeV at the LHC using molecular/four-quark coalescence model~\cite{ExHIC,ChoLee1}. The isospin assignment has been assumed to be  $I=0$. The calculations are detailed in Ref.~\cite{Abreu:2020ony}.  }
\label{Tab3}
\begin{center}
\begin{tabular}{c | c c }
\hline
\hline
State &  $  N_{X_{J}}  ^{  (4q)} (\tau_C) $ & $ N_{X_{J}}   ^{(Mol)}  (\tau_F) $    \\   
\hline
$J=0$  & $ 1.3 \times 10^{-3}$ & $ 4.5 \times 10^{-4}$
\\  
$J=1$  & $ 9.0 \times 10^{-4}$ & $ 7.2 \times 10^{-3}$
\\  
\hline
\hline
\end{tabular}
\end{center}
\end{table}

The present analysis takes into account the interactions of $X_{0,1}(2900)$ with the lightest and most abundant pseudoscalar meson $\pi$ constituting the surrounding hadronic medium. In light of this, we evaluate the contributions coming from the lowest-order Born diagrams for the reactions $X_J \pi \to \bar{D}^{\ast} K $ and $X_J \pi \to K^{\ast}  \bar{D} $, as well as the inverse processes. To calculate the respective thermally averaged cross sections used as input in the rate equation~(\ref{rateeq}), we employ three-body effective Lagrangians involving $\pi$, $K$, $D$, $K^*$  and $D^*$ mesons, i.e. ${\mathcal{L}}_{\pi D D^* }$ and ${\mathcal{L}}_{\pi K K^* } $~\cite{Abreu:2020ony}. 
The couplings incorporating the $X_{0,1}(2900)$ states have been written in order to yield the transition matrix elements $X_J \rightarrow \bar{D}^{\ast 0} K^{\ast 0} , D^{\ast -} K^{\ast +} $~\cite{Abreu:2020ony,Huang:2020ptc},   
\begin{eqnarray}\label{Lagr2}
{\mathcal{L}}_{X_0 \bar{D}^{\ast} K^{\ast} } &=& i g_{X_0 \bar{D}^{\ast} K^{\ast}} X_0 \bar{D}_\mu ^*  K^{\mu \ast} + H. c., 
\nonumber\\
{\mathcal{L}}_{X_1 \bar{D}^{\ast} K^{\ast} } &=& i g_{X_1 \bar{D}^{\ast} K^{\ast}}  X_1 ^{\nu} \bar{D}_\mu ^*  \overleftrightarrow{\partial_{\nu}}  K^{\mu \ast}.
\end{eqnarray}
In the expressions above, $X_0$ and $X_1$ denote respectively the $X_0(2900)$ and $X_1(2900)$ states according to Eq. (\ref{Xstates}). We emphasize that the isospin assignment has been assumed to be  $I=0$. 
The values of the coupling constants are chosen to be $g_{X_0 \bar{D}^{\ast} K^{\ast}} = 3.82 \substack{-0.16 \\ +0.11} \GeV$ and $g_{X_1 \bar{D}^{\ast} K^{\ast}} = 7.84 \substack{-0.30 \\ +1.00} \GeV$.

A last remark is concerning the characterization of the multiplicity of the particles in the hadronic medium. The total number of charm quarks $(N_c)$  in charm hadrons is assumed to be  conserved during the processes, i .e. $n_c(\tau) \times V(\tau) = N_c = 14$. This implies that the charm quark fugacity factor $\gamma _c $ in Eq.~(\ref{densities})  is time-dependent in order to keep $N_c$ constant. Besides, the total numbers of  pions and strange mesons at freeze-out were also based on 
Ref.~\cite{ExHIC}: $N_{\pi}(\tau_F) = 2410$ and $N_{K}(\tau_H)  = 134$. If we consider that the pions and strange mesons might be out of chemical equilibrium in the later part of the hadronic evolution, they would also have time dependent fugacities.


\begin{figure}[!ht]
\centering
\includegraphics[{width=10.0cm}]{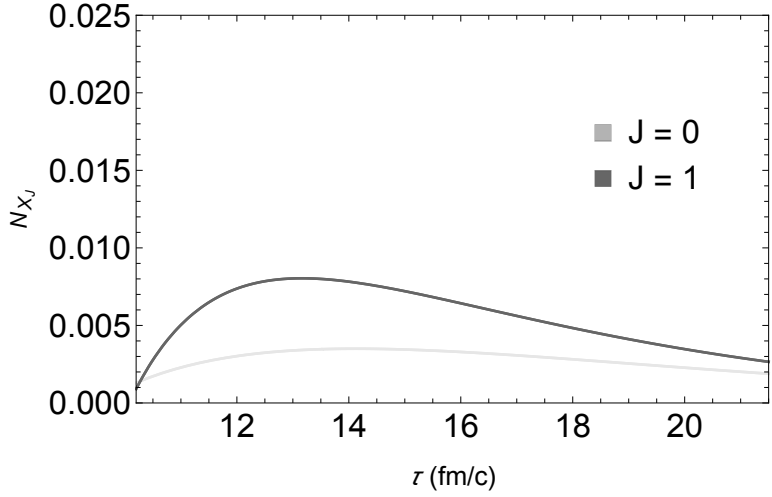}
\caption{Time evolution of the $N_{X_J}$ as a function of 
the proper time in central Pb-Pb collisions at $\sqrt{s_{NN}} = 5.02$ TeV at the LHC.    
Shaded bands represent the evolution of the number of  
$X_J$'s produced  at the end of the mixed phase 
calculated using tetraquark coalescence model.}
\label{TimeEvolXJ}
\end{figure}
%

We show  in Fig.~\ref{TimeEvolXJ} the time evolution of the $X_{0,1}(2900)$  abundances as a  
function of the proper time, taking $ N_{X_{0,1}}  ^{  (4q)}$ listed in Table~\ref{Tab3} as initial conditions. The outcomes suggest that the interactions of the $X_{0,1}$'s and the pions during the hadronic stage engender sizeable modifications: in the tetraquark coalescence model, the multiplicities suffer an increasing by a factor about 1.5 for $J=0$ and 3 for $J=1$. It is worthy mentioning that the terms in the last line of Eq.~(\ref{rateeq}), associated to the spontaneous decay/regeneration of $X_J$, play an important role in the change of $ N_{X_{0,1}} $ as the time goes by. 
Just for the sake of comparison: if $\Gamma _{X_0, X_1}$ were assumed to be zero, in the case $J=0$ the number of  $X_0$'s  throughout the hadron gas phase would be almost constant; for $J=1 $ the increasing of the multiplicity would be by a factor  about $25\% $.

\begin{figure}[!ht]
\includegraphics[{width=10.0cm}]{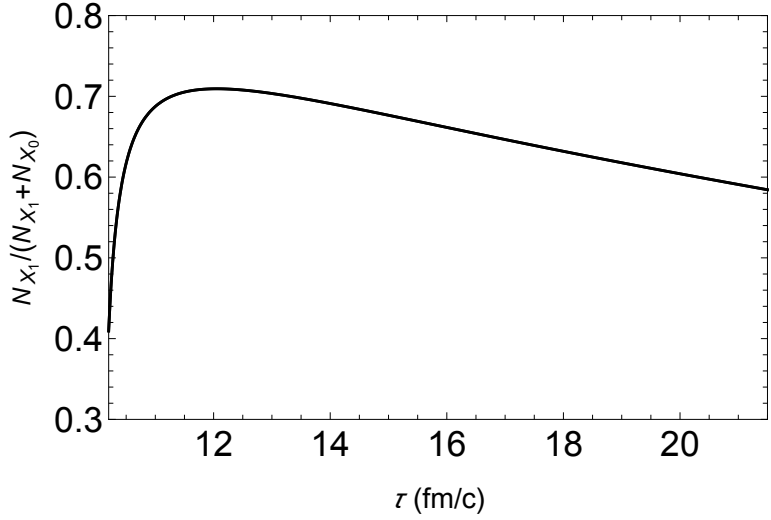}
\centering
\caption{Time evolution of the ratio of the $X_1$ abundance to the sum of the  $X_0$ and $X_1$ abundances, as a function of 
the proper time in central Pb-Pb collisions at $\sqrt{s_{NN}} = 5$ TeV at the LHC, using tetraquark coalescence model.}
\label{TimeEvolXJJ0OVERJ0PLUSJ1}
\end{figure}

In Fig.~\ref{TimeEvolXJJ0OVERJ0PLUSJ1} is plotted the evolution of the ratio of the $X_1$ abundance to the sum of the  $X_0$ and $X_1$ abundances. It can be seen that the ratio experiences an sudden increase from $41\%$ to $71\%$, and a further reduction up to $58\%$ in the end. In other words, tetraquark coalescence model gives the $X_1$-state with a slight higher multiplicity at kinetic freeze-out.

%

We wish to emphasize that the estimations of time evolution of $X_{0,1}(2900)$ abundances reported above are based on initial conditions from the tetraquark coalescence model.  However, these findings deserve a comparison with the other possible formation mechanism dominant at the end of the hadronic phase, i.e. the hadron coalescence. Comparing the values $ N_{X_{J}}   ^{(Mol)}  (\tau_F) $  listed in Table~\ref{Tab3} with the results in Fig.~\ref{TimeEvolXJ}  at kinetic freeze-out, we see that the ratio is about 4 times smaller for $J=0$  and 2.5 greater for $J=1$. This means that at the end of the hadron gas phase the  yield of $X_0$ is mainly due to the compact tetraquark configuration, whereas for $X_1$ the most leading production comes from the hadronic molecular state.

\section{Concluding remarks}

\label{Conclusions}

We have discussed in this work how the production and dissociation of hadron states can be affected by reactions during the expansion of hadronic matter in a heavy-ion collision environment. The main idea presented, based on previous works, is that these interactions might be relevant in some cases, and this analysis appears as an interesting tool to discriminate the intrinsic structure of the recently observed exotic hadronic states. 
A case study has been revisited here: the time evolution of the abundances of the $X_{0,1}(2900)$ states. This has been done by solving the rate equation, and the findings showed that they are affected during the expansion of the hadronic matter. Considering the $X_J$ as a tetraquark state produced via quark coalescence mechanism from the QGP phase, in which $X_0$ is a relative $S$-wave and $X_1$ a $P$-wave, the results suggest an increasing by a factor about 1.5 for $J=0$ and 3 for $J=1$ at kinetic freeze-out. Also, the $X_1$-state has a slight higher multiplicity at the end of hadronic phase.

The comparison of the $X_{0,1} (2900)$ multiplicities obtained from the hadron coalescence ($J=0$ as a $S$-wave and $J=1$ as a $P$-wave) and tetraquark coalescence at kinetic freeze-out, allowed to infer that the yield of $X_0$ is mainly due to the compact tetraquark configuration, whereas for $X_1$ the most leading production comes from the hadronic molecular state. 
Therefore, if a vertex detector cumulate by about $10^3$ charmed mesons, based on  our findings a few $X_0 (2900)$ and $X_1 (2900)$ are expected to be yielded if they are $S$-wave tetraquark state and $P$-wave hadronic molecular state for $J=0$ and $J=1$, respectively. Improvements on this effective approach and on the hydrodynamical model are under development in order to reach an even better phenomenological description.


\section*{Acknowledgement}


The author would like to thank the Brazilian funding agencies for their financial support: CNPq (contracts 308088/2017-4 and 400546/2016-7) and FAPESB (contract INT0007/2016).


\end{document}